\documentclass[nofootinbib,twocolumn,showpacs,preprintnumbers,amsmath,amssymb]{revtex4}

\usepackage{graphicx}
\usepackage{dcolumn}
\usepackage{bm}

\begin{document}

\title{Testing a Dilaton Gravity Model using Nucleosynthesis}

\author{S. Boran}
\email{borans@itu.edu.tr}
\author{E. O. Kahya}
\email{eokahya@itu.edu.tr}
\affiliation{Department of Physics, {\.I}stanbul Technical University, Maslak 34469 Istanbul, Turkey}

\date{\today}
\begin{abstract}
Big Bang Nucleosynthesis (BBN) offers one of the most strict evidences for 
the $\Lambda$-CDM cosmology at present, as well as the Cosmic Microwave Background
(CMB) radiation. In this work, our main aim is to present the outcomes 
of our calculations related to primordial abundances of light elements, 
in the context of higher dimensional steady-state universe model in 
the dilaton gravity. Our results show that abundances of light elements 
(primordial D, $^{3}$He, $^{4}$He, T, $^{7}$Li) are significantly different 
for some cases, and a comparison is given between a particular dilaton gravity 
model and $\Lambda$-CDM in the light of the astrophysical observations.

\end{abstract}

\maketitle

\section{Introduction}\label{introduction}
The current expansion of the universe is a crucial evidence for the big bang 
cosmology model. It predicts the chemical abundances of primordial elements as 
results of nuclear reactions which began seconds after the big bang and 
continued for the next several minutes. With the help of inflation, one can 
consistently solve the well-known problems of the standard model; such as 
the observed spatial homogeneity, isotropy, and flatness of the universe \cite{ADL}. 

There are still many unsolved puzzles of this model, such as the origin of 
dark matter and dark energy, cosmological constant problem, cosmic coincidence 
problem and the exact form of the inflation potential etc. \cite{PR, SP} On the other 
hand, there are many models which claim solutions to these problems by modifying 
Einstein's general relativity. Quintessence, k-essence, phantom, quintom and other 
phenomenological models are just few examples of alternate gravity models that 
offer a solution to the dark energy problem \cite{DM}. And also there are alternative 
gravity theories that suggest using extra fields (scalar-tensor, etc.) and higher 
dimensions (Kaluze-Klein, Randall-Sundrum) arising from string theory at the low 
energy limit \cite{RS}.

In this ocean of models, we would like to consider an observable consequence 
(modified abundances of light elements) of a new model, a Higher Dimensional 
Dilaton Gravity Theory of Steady-State Cosmological (HDGS) model in the context 
of string theory. We need to highlight that the original steady state model \cite{BG, H} 
is unfavorable compared to the standard big bang scenario. But our motivation 
in this work is to suggest a test of a specific higher dimensional dilaton gravity 
model which effectively mimics the standard FRW model with the modified Hubble constant. 
Hence an immediate consequence would be the modification of nucleosynthesis. 

This modification was investigated in \cite{OT} and it was claimed that this model 
gives a better estimate for the primordial $^{4}$He abundance compared to 
the Standard Big Bang Nucleosynthesis (SBBN) by choosing the number of dimensions 
appropriately. In this work, to further test their strong claim, we calculated 
the abundances of the primordial D, $^{3}$He, $^{4}$He, T, $^{7}$Li 
in the context of this nonstandard (HDGS) model and compared it with the predictions 
of SBBN and the astrophysical observations.

On the other hand, at high energies the quantum gravitational corrections 
will start to play an important role. Quantum corrections will modify 
the dilaton gravity models as well and therefore  change the whole form of 
this model via action \cite{NO1, NO2, NO3, NO4}. One would naturally expect 
to see quantum effects during the very early universe such as primordial 
inflationary stage. During inflation, quantum loop effects may lead to 
very small \cite{W1, W2} but possibly observable corrections to power spectrum 
\cite{SS1, SS2, MM, GSS, EV, EW}. Therefore one might describe the interactions 
with effective field theories of inflation \cite{CS}. But in this work, we are 
mainly interested in the consequence of a geometrical constraint: 3+$n$ dimensional 
universe having a constant volume, leading to a modified Hubble parameter during 
a later stage, nucleosynthesis, where quantum gravitational corrections are negligible. 

The paper is organized as follows: In $\S$\ref{dg}, dynamics of this particular 
dilaton gravity model is summarized. In $\S$\ref{ndg}, nucleosynthesis in the context 
of this model is analyzed. In $\S$\ref{discussion}, the results obtained from our 
calculations for this model and the predictions of SBBN with the help of 
Plank Satellite data \cite{PS} are compared with the astrophysical observations.
\section{Dynamics of HDGS, a Dilaton Gravity Model}\label{dg}
In this section, we briefly summarize the dynamics of a particular type of dilaton
gravity models that proposed in \cite{OT}. The idea is introducing a higher 
dimensional dilaton gravity action of steady state cosmology (HDGS) in the string frame. 
Therefore the evolution of the internal n-dimensional space results into an evolution 
of the observed universe to keep the whole system in a steady state. Due to this 
constraint choosing particular values for some parameters, such as the number of extra 
dimensions, leads to possibly observable effects in our universe. Let us start with 
the action, which stems from the low-energy effective string theory,
\begin{equation}\label{stringaction}
S = \int_{M} d^{1+3+n} x \sqrt{|g|} e^{-2 \phi} (R + 4\omega \partial_{\mu} \phi \partial^{\mu} \phi + U(\phi)),
\end{equation}
where $R$ is the curvature scalar, $M$ stands for manifold, $n$ corresponds 
to extra dimensions, $|g|$ is the determinant of $g_{\mu\nu}$ metric tensor, $\phi$ 
is the dilaton field taken as space independent real function of time and $\omega$ 
is an arbitrary coupling constant. $U(\phi)=U_{0}e^{\lambda\phi}$ is a real smooth 
function of the dilaton field and corresponds to the dilaton self interaction potential 
and both $U_{0}$ and $\lambda$ are real parameters. Two interesting cases that is 
worth mentioning are $\omega=1$ and n=6 corresponding to anomaly-free superstring 
theory and $\omega=1$ with n=22 corresponds to bosonic string theory. The metric 
is given by 
\begin{equation}\label{met}
ds^{2}=-dt^{2}+a^{2}(t)(dx^{2}+dy^{2}+dz^{2})+s^{2}(t)(d\theta_{1}^{2}+...+d\theta_{n}^{2}).
\end{equation}
Here t is the cosmic time, $(x,y,z)$ are the cartesian coordinates of the 3-dimensional 
flat space, basically the observed universe. The coordinates, $\theta$ are $n$-dimensional, 
compact (torodial) internal space coordinates (this represents space that cannot be observed 
directly and locally today.). While $a(t)$ denotes the scale factor of 3-dimensional 
external space, $s(t)$ is the scale factor of $n$-dimensional internal space. 

This model has the following key properties: 

(i) The $(3+n)$-dimensional universe has a constant volume, that is $V=a^{3}s^{n}=V_{0}$, 
hence steady state. But the internal and external spaces are dynamical. (ii) The energy density 
is constant in the higher dimensional universe. (iii) There is no higher dimensional matter 
source other than the dilaton field in the action.

If the scalar field is redefined as $\beta = e^{\lambda \phi}$, the relation 
between the scalar field and the scale factor of the external space turns out to be
\begin{equation}
(\frac{a^{'}}{a})^{2} = \frac{n}{3(3+n)}\frac{2\omega\varepsilon}{\beta ^{2}},
\end{equation}
where $\varepsilon$ is a constant of integration. Here, prime denotes 
derivative with respect to the ordinary time. Imposing the constant volume condition gives,
\begin{equation}
a=a_{0}e^{\pm\frac{1}{3}\sqrt{\frac{3n}{(3+n)}} \sqrt{2\omega \varepsilon} \int dt \frac{1}{\beta}}
\end{equation}
and 
\begin{equation}
s=s_{0}e^{\mp\frac{1}{n}\sqrt{\frac{3n}{(3+n)}} \sqrt{2\omega \varepsilon} \int dt \frac{1}{\beta}}
\end{equation}
where $a_{0}$ and $s_{0}$ correspond the integration constants.
Therefore the modified Hubble parameter of the external space is obtained 
as follows 
\begin{equation}\label{mha}
H_{a}\equiv \frac{a^{'}}{a} = \pm \frac{1}{3} \sqrt{\frac{3n}{3+n}}\frac{\sqrt{2\omega\varepsilon}}{\beta}.
\end{equation}
Here the physically relevant case is the solution for expanding external space with $H_{a} > 0$. 
The deceleration parameter for the external space is given by 
\begin{equation}\label{mda}
q_{a}\equiv -\frac{a^{''}a}{a^{'2}} = -1 \pm 3 \sqrt{\frac{3+n}{3n}}\frac{\beta^{'}}{\sqrt{2\omega\varepsilon}}.
\end{equation}
In the case of $\varepsilon\neq0$ and $U_{0}\neq0$, and with the choice of appropriate initial 
conditions, it turns out that \cite{OT} the early time modified deceleration parameter is given by
\begin{equation}\label{qaz}
q \rightarrow \,  3\sqrt{\frac{3+n}{3n}}\frac{1}{\sqrt{\omega}}-1.
\end{equation}
\section{Nucleosynthesis in HDGS}\label{ndg}
We are interested in how abundances of light elements would change in the context 
of this model. Specifically we would like to consider the ratio of the modified 
expansion rate to standard expansion rate during the early radiation dominant 
epoch. This ratio is given by,
\begin{equation}\label{S}
S \equiv \frac{{H_a}}{H_{SBBN}} = \frac{1+q_{SBBN}}{1+{q_a}}.
\end{equation}
This is true since deceleration parameter stays almost constant during primordial 
nucleosynthesis. The value of the deceleration parameter for standard BBN is $q_{SBBN}=1$. 
Since $q_a$ is given by (\ref{qaz}), the so-called standard expansion factor, S 
can be expressed in terms of $\omega$ and $n$ as 
\begin{equation}\label{mS}
S = \frac{2}{3}(\sqrt{\frac{3\omega n}{3+n}}).
\end{equation}

If $S\neq1$ is taken, it denotes nonstandard expansion factor. This kind of 
modification might also arise due to additional light particles such as neutrinos 
which would make the ratio to be, $\tilde{H}/H_{SBBN}=[1+\frac{7}{43}(N_{\nu} - 3)]^{\frac{1}{2}}$. 
In this context of the dilaton gravity model that we mentioned it is also going to 
occur due to a modification of general relativity. We are interested in the case 
where $N_{\nu}=3$ and therefore the value of $(S-1)$ will come only from the modification 
of general relativity. 

The primordial abundances of the light elements (primordial D, $^{3}$He, 
$^{4}$He, $^{7}$Li, T) depend on the baryon density and the expansion rate 
of the universe \cite{VSS,GS}. The baryon density parameter \cite{GS} is given by 
\begin{equation}\label{etaB}
\eta_{10} \equiv 10^{10} \eta_{B} \equiv 10^{10} \frac{n_{B}}{n_{\gamma}} = 273.9 \Omega_{B}h^{2},
\end{equation}
where $\eta_{B}$ gives the baryon to photon ratio, $\Omega_{B}$ is 
dimensionless current critical cosmological density parameter for baryons and 
$h=h_{100}\equiv \frac{H_{0}}{100kms^{-1}Mpc^{-1}}$ with $H_{0}$ being 
the present value of the Hubble parameter. Any modification of the expansion 
rate would change the time when neutrons freeze out, which will in turn determine 
the final abundance of Helium-4 as well as all of the other light elements.  

In the following subsections, we will analyze nucleosynthesis due to a modification 
of the expansion rate in the context of HDGS models. We will express the primordial 
nuclear abundances of light nuclei in terms of two parameters of HDGS models; number 
of extra dimensions $n$, and coupling constant $\omega$. Particularly, we will be 
interested in the case of$\omega=1$, where $n=6$ and $n=22$ correspond to anomaly-free 
superstring and bosonic string theory, respectively. 
\subsection{$^{4}$He abundance in HDGS models}
The two body reaction chains of light elements, which include Deuterium (D), 
Tritium (T) and Helium-3 ($^{3}$He) to produce Helium-4 ($^{4}$He), are more 
efficient than four body reactions of neutrons and protons. The first step is 
producing D from $n+p\rightarrow D + \gamma$. After that D is converted into 
$^{3}$He and T; 
\begin{equation}\label{fbody}  D+D \rightarrow \, ^{3}He+n \;\;\;\;\;\;\;\;  D+D \rightarrow \, T+p
\end{equation}
and finally, $^{4}$He is produced from D combining with T and $^{3}$He;
\begin{equation}\label{sbody}  T+D \rightarrow \, ^{4}He+n \;\;\;\;\;\;\;\;  ^{3}He+D \rightarrow \, ^{4}He+p.
\end{equation}

In order to get precise estimates for abundances of light elements, one should solve 
non-linear differential equations of the nuclear reaction network. This problem can be 
studied numerically and the modern methods are based on Wagoner \cite{Wagoner}code 
and its updated version by Kawano \cite{Kawano}. The next step is getting a best fit 
to a numerical work to see how various abundances depend on $\eta_{10}$ and other 
parameters such as number of extra neutrinos etc. Another venue is applying 
semi-analytical methods; where one of the earliest work was done by Esmailzadeh 
\textit{et al.} \cite{Esmail} using method of fixed points. 

In this work we would like to use, if there exists, the best-fit expressions for 
certain elements. If there is none in the literature for a certain element, then we 
will use a semi-analytical approach that is based on a simple assumption, which is 
the nuclear reaction network obeying in a quasi-equilibrium state. In this state a 
basically one assumes that ``the total flux coming into each corresponding reservoir 
must be equal to the outgoing flux'' \cite{MUKH}. 

A simple way of estimating of $^{4}$He abundance\footnote{In general 
abundance by weight is related to the ratio of number density of a particular element 
to the number density of all nucleons(including the ones in complex nuclei), $X_A \equiv A n_A / n_N$, 
where A is the mass number of a particular element, e.g. $A=4$ for Helium.} is the following: 
multiply the abundance of neutrons by two at the time when the deuterium 
bottleneck opens up. Here we will refer to the best fit expression for $^{4}$He abundance 
that includes the case of modified expansion rate \cite{GK, GS7}:
\begin{equation}\label{Yp}
Y_{p} = 0.2485 \pm 0.0006 + 0.0016((\eta_{10} - 6) + 100(S-1)),
\end{equation}
where p stands for the primordial abundance. We will take $\eta_{10} \simeq 6$ \cite{eta10} 
from here on. The SBBN value, $S=1$, becomes $Y_{p}^{SBBN} = 0.2485 \pm 0.0006$. Using 
equation (\ref{mS}),for the case of HDGS models that we are interested in, one can gets 
the following expression for $^{4}$He abundance in terms of $\omega$ and $n$ as \cite{OT} 
\begin{equation}\label{Ypstring}
Y_{p}= 0.2485 \pm 0.0006 + 0.16 (-1 + \frac{2}{3}\sqrt{\frac{3\omega n}{3+n}}).
\end{equation}
In the case of $\omega=1$, the predicted $Y_{p}$ values are obtained 
as  $Y_{p}=0.2393 \pm 0.0006$ and $Y_{p}=0.2618 \pm 0.0006$, for $n=6$ and $n=22$, 
respectively.

From the observational point of view, the $^{4}$He primordial abundance, $Y_{p}$ 
is determined from recombination of lines of the H II from blue compact 
galaxies (BCGs) \cite{Pagel}. The observational results of the $^{4}$He
abundances are given by $Y_{p} = 0.2565 \pm 0.0060$ \cite{52} and 
$Y_{p} = 0.2561 \pm 0.0108$ \cite{53}.
\subsection{Abundances of other light elements in HDGS models}
\subsubsection{Deuterium abundance} 
Deuterium is produced by $ p + n \rightarrow D + \gamma$ and used in four 
types of reactions (\ref{fbody}), (\ref{sbody}). Therefore one would expect to 
solve either numerically or analytically the equations for this nuclear reaction 
network and get the expression for deuterium abundance, $X_{D} \equiv 2n_{D}/n_N$, 
where $n_{D}$ and $n_{N}$ are the number densities of deuterium and all nucleons, 
respectively. 

In literature, instead of abundances of elements, their abundances relative to 
hydrogen are given. To see why let us look at how deuterium is determined. 
The absorbed this primordial element has more space in the wings of the observed 
quasar absorption-line systems (QAS) \cite{23,24,25,26,27} than the absorbed 
hydrogen at high redshifts (z) and/or at low metallicity (Z). Also, the observation 
of the multicomponent velocities of these absorbed elements is very significant 
in order to determine the abundance of deuterium. Therefore the $(\frac{D}{H})_{p}$ 
ratio is more meaningful, and often known as interstellar medium measurement 
for deuterium abundance. This ratio can be expressed in terms of the abundance 
by weight of the deuterium as
\begin{equation}\label{ryDp} 
y_{Dp} \equiv 10^{5} (\frac{n_D}{n_H})_{p} = 10^{5} (\frac{13}{24}X_{Dp}).
\end{equation}
The factor, $\frac{13}{24}$ comes from the fact that mass number of deuterium is 2 
and hydrogen number density is equal to $\frac{12}{13}$ of all the nucleons in the 
universe, i.e. 75\% by weight.

Let us start with the semi-analytical expression for the abundance of deuterium 
to calculate (\ref{ryDp}). Using the quasi-equilibrium condition one can get \cite{MUKH}
\begin{equation}\label{XDfquasi}
X_{Dp} \simeq \frac{2R}{exp(A\eta_{10}) - 1} \simeq 4.87 \times 10^{-5},
\end{equation}
where $R\simeq 2\cdot 10^{-5}$ \cite{MUKH}, $\eta_{10}\simeq 6$ and $A\simeq0.1$. 
Here the coefficients $R$ and $A$ are related to experimental values of nuclear reaction 
rates\footnote{We assume that the nuclear interaction rates are independent of extra dimensions.},\footnote{There are no matter sources in higher dimensions and HDGS is a Kaluza-Klein type model rather than a brane world cosmology one.} of deuterium at temperature of order 0.08$ MeV$. Putting this value in (\ref{ryDp}) 
gives $y_{Dp}^{SBBN} = 2.63$.

Let us now use a more precise expression for deuterium abundance \cite{GS} based on 
a numerical best fit:
\begin{equation}\label{yDp}
y_{Dp} = 2.60(1 \pm 0.06)(\frac{6}{\eta_{10}-6(S-1)})^{1.6}.
\end{equation}
From this expression one can get the SBBN value of $y_{Dp}$ (for $S=1$ and $\eta_{10}\simeq 6$) 
as $y_{Dp}^{SBBN} = 2.60 \pm 0.16$. Comparing this number with the one from semi-analytical 
method, $y_{Dp}^{SBBN} = 2.63$, we can safely assume a quasi equilibrium condition, if necessary. 

By using equation (\ref{mS}), one can express $y_{Dp}$ for HDGS models as,
\begin{equation}\label{yDpstring}
y_{Dp}= 2.60(1 \pm 0.06)(\frac{6}{\eta_{10}-6(-1 + \frac{2}{3}\sqrt{\frac{3\omega n}{3+n}})})^{1.6}.
\end{equation}
Taking $\eta_{10}\simeq6$, the predicted values of $y_{Dp}$ are obtained as 
$y_{Dp} = 2.38 \pm 0.16$ and $y_{Dp}=2.99 \pm 0.16$ for $n=6$ and $n=22$, 
respectively for $\omega=1$ model.

Finally the observational results are $y_{Dp} = 2.87 \pm 0.22$ \cite{I} and 
$y_{Dp} = 2.54 \pm 0.05$ \cite{27}.
\subsubsection{Helium-3 abundance} 
The relevant nuclear reactions that involve $^{3}$He are:
\begin{eqnarray}  D+D \rightarrow \, ^{3}He+n \;\;\;\;\;\;\;\;\;\;  D+p \rightarrow \, ^{3}He+\gamma
\\           ^{3}He+n \rightarrow \, T+p \;\;\;\;\;\;\;\;\;\;  ^{3}He+D \rightarrow \, ^{4}He+p .
\end{eqnarray}

The quantity used in the literature to describe $^{3}$He is
\begin{equation}
\label{ryHp} y_{3} \equiv 10^{5} (\frac{n_{^{3}He}}{n_H}) = 10^{5} (\frac{13}{36}X_{^{3}He}).
\end{equation} 
Making a quasi-equilibrium approximation for $^{3}$He abundance we can express the $^{3}$He 
abundance in terms of deuterium abundance after using the experimental values for the ratios 
of the related nuclear reaction rates \cite{MUKH};
\begin{equation}\label{he3quasi}
X_{^{3}He} \simeq \frac{0.2\cdot X_{D} + 10^{-5}}{1+ 4 \times 10^{3} X_{D}}.
\end{equation}
From this equation we can see that $^{3}$He abundance is not as sensitive as deuterium 
since a change in deuterium abundance would change both parts of the ratio. 
One can also see this from the weaker dependence of $y_3$ on $\eta_{10}$, compared to 
$y_{Dp}$, for SBBN best fit expression \cite{GSp}.
\begin{equation}\label{yDp}
y_{3} = 3.1(1 \pm 0.03) \eta_{10}^{-0.6}.
\end{equation}  
Therefore $^{3}$He abundance is not a good indicator of a modification of SBBN 
due to HDGS models.
\subsubsection{Tritium abundance}
Using the quasi-equilibrium condition for tritium, 
$X_{T}^{\phantom{\eta}f}$ \cite{MUKH} is obtained as 
\begin{equation}\label{Tquasi}
X_{T}^{\phantom{\eta}f} \simeq (0.015 + 3\cdot 10^{2} X_{^{3}He}^{\phantom{\eta\beta}f})X_{D}^{\phantom{\eta}f}.
\end{equation}
It is clear from this expression that the value of tritium abundance will be 
as sensitive as deuterium abundance to any modification of the expansion rate. But the 
magnitude of tritium abundance is two orders of magnitude smaller than both 
Deuterium and Helium-3. Therefore observationally it is not very feasible 
but it should be kept in mind that it can be used to test for consistency in the future 
experiments.  
\subsubsection{Lithium-7 abundance} 
Finally we would like to investigate the effects 
of modified expansion rate to lithium abundance. The $^{7}$Li abundance is given by
\begin{equation}\label{ryLip}
y_{Lip}\equiv 10^{10}(\frac{n_{Li}}{n_{H}})_{p}.
\end{equation}
One might think that its smallness would make it irrelevant for observational purposes. 
But, it can actually be measured in the atmospheres of metal-poor stars in the stellar 
halo of Milky-way. The puzzling part is that given the $\eta_{10}$ parameter, which almost 
fits all the other elements successfully, results into a discrepancy for lithium. 
The ratio of the expected SBBN value of Lithium-7 abundance to the observed one is between 
2.4-4.3 \cite{BDF}. Therefore it should be interesting to check if this HDGS models 
offer any solution to the \textit{lithium problem}.

The best fit expression to the numerical BBN data of the $y_{Lip}$ is given in \cite{GS} as
\begin{equation}\label{yLip}
y_{Lip} = 4.82(1 \pm 0.10)(\frac{\eta_{10}-3(S-1)}{6})^{2}.
\end{equation}
Taking $S=1$ and $\eta_{10} \simeq 6$, 
the SBBN value of Lithium-7 abundance is found as $y_{Lip}^{SBBN} = 4.82 \pm 0.48$.
In terms of $\omega$ and $n$, the modified form of the equation (\ref{yLip}) becomes 
\begin{equation}\label{yListring}
y_{Lip} = 4.82(1 \pm 0.10)(\eta_{10}-3(-1 + \frac{2}{3}\sqrt{\frac{3\omega n}{3+n}}))^{2}.
\end{equation}
By using the equation (\ref{yListring}), the predicted $y_{Lip}$ values are found as 
$y_{Lip} = 5.10 \pm 0.51$ for $n=6$ and $y_{Lip} = 4.43 \pm 0.44$ for $n=22$, 
for the case of $\omega=1$.   
\section{Discussion}\label{discussion}
We have shown in this work that one gets a considerable modification to the primordial 
abundances of light elements in the case of a higher dimensional steady state universe in 
dilaton gravity\footnote{There are other ways to modify BBN based on scalar-tensor theories, 
for details see \cite{DP, AKJE, CKMJ} and references therein.}. Although there is a huge class of models that 
one can consider, with two free parameters $\omega$ (dilaton coupling constant) and $n$ 
(number of internal dimensions), we focused on two interesting cases where $\omega=1$ and 
$n=6$, $n=22$ that corresponds to anomaly-free superstring and bosonic string theory, 
respectively. 

The main idea behind the calculation is modifying the expansion rate during the 
nucleosynthesis to get different abundances for light elements. One can think the 
modification as being similar to adding more relativistic particles, such as 
extra neutrinos, into the standard big bang model. When Hubble parameter gets 
modified all the nucleosynthesis will get modified as well. The question is the 
following: Is this modification large enough to observe and if it is then 
is it compatible with the data?

To answer these questions one should analyze how the nuclear reactions get modified with 
the modification of the expansion rate. It is well-known that the complete analysis of 
the nuclear reactions governing the primordial abundances of light elements can be done 
using numerical methods. We used the results of the previous works, where we can, which 
were obtained by getting best fit expressions to numerical data related to the abundances 
of these elements. And if there are no known best-fit expressions in the literature 
we proceeded our analysis based on semi-analytical methods.
   
The primordial abundance of Helium-4 was already studied in the context of these models. 
It was pointed out that $n=22$ case is more compatible with the Helium-4 data compared to 
the standard big bang scenario. We made a more extensive analysis of other light elements 
and checked the compatibility of this model with astrophysical observables. The results are 
summarized in TABLE I.

One can clearly see from the TABLE I that $\omega=1$ and $n=6$ dilaton gravity model is 
incompatible with Helium-4 data and is incompatible with Deuterium as well. Helium-4 data 
favoured the case of $\omega=1$ and $n=22$ compared to SBBN, as was noted. In the case of
deuterium earlier measurements favour (with almost being inside the error bars) 
dilaton gravity model whereas the more recent measurements rule them out and point 
towards SBBN. Therefore it is fair to say that one needs more observations and data 
analysis to see which model is favoured.  

We also showed that Helium-3 and Tritium abundances are not very convenient to see 
a modification of the standard model, in the context of the dilaton gravity model 
considered here. And for the case of Lithium-7 one gets almost a ten percent decrease 
for the expected abundance, compared to SBBN, but it still is far from explaining 
the observed abundance. So, these models do not offer a solution to the \textit{lithium 
problem}, therefore the existence of this problem still preserves its place in the 
literature and leaves an open window to new physics.
\begin{widetext}

\begin{table} 
\begin{center}
\caption{The abundances He-4, Deuterium and Li-7 for different models}
\begin{ruledtabular}
\begin{tabular}{lccc}
Models and Data / Abundances:&$Y_{p}$&$y_{Dp}$&$y_{Lip}$ \\
\hline
SBBN model:\;\;\;\;\;\;\;&$0.2485 \pm 0.0006$&$2.60 \pm 0.16$&$4.82 \pm 0.48$ \\
n=6 Dilaton gravity model: \;\;\;\;\;\;\;&$0.2393 \pm 0.0006$&$2.38 \pm 0.16$&$5.10 \pm 0.51$ \\
n=22 Dilaton gravity model:\;\;\;\;\;\;\;&$0.2618 \pm 0.0006$&$2.99 \pm 0.16$&$4.43 \pm 0.44$ \\ 
Observational data:&\;\;\;\;\;$0.2561 \pm 0.0108$\cite{53}&\;\;\;\;\;$2.88 \pm 0.22$\cite{I}&\;\;\;\;\;$1.1 - 1.5$\cite{Asp} \\
                   &\;\;\;\;\;$0.2565 \pm 0.0060$\cite{52}&\;\;\;\;\;$2.54 \pm 0.05$\cite{27}&\;\;\;\;\;$1.23^{+0.68}_{-0.32}$\cite{Ryan} \\
\end{tabular}
\end{ruledtabular}
\end{center}\label{t1}
\end{table}

\end{widetext}
\begin{acknowledgments}
We would like to thank \"{O}zg\"{u}r Akarsu, Ali Kaya and Subir Sarkar for 
helpful discussions. This work was supported by T\"{U}B{\.I}TAK-1001 Grant No.112T817.
\end{acknowledgments}

\end{document}